\documentclass[aps,floats,amssymb,twocolumn,prb]{revtex4}
\usepackage{calc}
\usepackage{psfrag}
\usepackage{graphicx}
\usepackage{color}

\begin{document}

\title{Paired states in half-filled Landau levels}

\author{M. V. Milovanovi\'c$^{1,2}$}
\affiliation{$^1$ Scientific Computing Laboratory, Center for the Study of Complex Systems, Institute of Physics Belgrade, University of Belgrade, Pregrevica 118, 11080 Belgrade, Serbia}
\affiliation{$^2$ Kavli Institute for Theoretical Physics, Santa Barbara, CA 93106, USA}

\begin{abstract}
We discuss monolayer and bilayer quantum Hall systems in which each layer is a half-filled Landau level (LL) system. In the mean field approximation of the Son's formalism there is a common pairing structure that underlines the possibilities for paired ground states in both systems. We argue that the particle-hole (PH) Pfaffian state in the (particle-hole symmetric) half-filled LL of a monolayer, and analogous state in the PH symmetric bilayer
(in which each layer is half-filled LL) can be considered as {\em critical states} i.e. states that cannot describe a phase under PH symmetry. We point out that the inclusion of a PH symmetry breaking (like LL mixing) may stabilize the PH Pfaffian in a monolayer. In the bilayer case, in numerical experiments on a sphere, by choosing the PH symmetric shift, we can stabilize the interlayer correlated (111) excitonic state or critical state, for any distance between the layers, but in general, with no bias, the evolution of the bilayer includes other phases.
\end{abstract}

\maketitle
\section{Introduction}

The fractional quantum Hall effect (FQHE) is a strongly correlated phenomenon  two dimensions (2D), in which a system of electrons has Hall conductance that takes fractional values for intervals of magnetic field strength or system density. The effect can be explained by studying the system at  a particular filling factor i.e. fraction associated with the effect.  The filling factor is the ratio between the number of electrons and the number of available flux quanta in a Landau level (LL). This commensuration, associated with an effective projection to a fixed LL, leads to exotic phenomena, which include pairing of electrons in the presence of repulsive interactions. Thus (classically speaking)
 phase space constraints may lead to a BCS pairing physics (without superconductivity). The prime example, introduced in [\onlinecite{mr}], of this phenomenon is in the FQHE system of spinless electrons at filling fraction 1/2,
with a special (BCS) pairing function that is called Pfaffian. The Pfaffian construction supports quasiparticles with non-Abelian statistics, and may find application in future quantum computers [\onlinecite{sns}].

At least theoretically, we may envision a situation when the description of a FQHE system is confined to a fixed LL (i.e. all available states are in a fixed LL). At the exact filling of a fixed LL equal to 1/2, we have a symmetry between particles and holes i.e. the PH symmetry. The Pfaffian construction as well its PH conjugate counterpart, anti-Pfaffian [\onlinecite{lhr,nf}], do not possess this symmetry. The question that may be asked is whether a Pfaffian-like construction that respects the PH symmetry, so-called PH Pfaffian, exists in a fixed LL. Recently, a theoretical proposal [\onlinecite{son}] was put forward concerning  a non-quantized (non-paired) FQHE state - the Fermi-liquid-like state of dressed electrons i.e. composite fermions (CFs) at filling factor 1/2 [\onlinecite{hlr}]. According to the proposal the system at filling factor 1/2, that is also PH symmetric, behaves as a system of (relativistic - two component) Dirac CFs at finite chemical potential (fixed by the strength of external magnetic field). With this new insight into the nature of the (non-paired) metallic state of CFs at filling 1/2 (PH symmetric half-filled LL), came a proposal for a PH Pfaffian state (a paired state of Dirac CFs) [\onlinecite{son}].

Interestingly enough, numerical experiments [\onlinecite{rh,pjds,dh,ph,pp}], at a half-filled LL, testify in favor of superpositions of Pfafian and anti-Pfaffian (if no spontaneous breaking of the PH symmetry occurs), in the PH symmetric case. By taking into account previous model wave function ansatzes for PH Pfaffian [\onlinecite{tj,mvmtj}], and using the mean field treatment of the Son's formalism [\onlinecite{son}], we argue that in the strict PH symmetric circumstances, PH Pfaffian is a critical state, and cannot describe a stable FQHE phase. But if we consider, for example LL mixing, and break PH symmetry by a mass term in the Son's formalism, we can stabilize PH Pfaffian as a topological phase.

The second system that we will consider is the bilayer at the total filling one, i.e. each layer is represented by a half-filled (lowest) LL in a PH symmetric set-up. When the distance between the layers, $d$, is small, $ d << l_B$; $l_B$ is the magnetic length, and the inter- and intra-correlations are about the same. At this special filling factor, they lead to an exceptional intercorrelated state [\onlinecite{wz,moon}]: this state supports a counterflow superfluidity and a Goldstone mode
[\onlinecite{eis}]. On the other hand at $d \rightarrow \infty$, we have two decoupled layers, where each one represents the Fermi-liquid-like state of (Dirac) CFs in  the lowest LL.

The evolution of the bilayer with changing distance is the subject of many analytical and numerical [\onlinecite{th1,th2,th3,th4,th5,th6,th7,th8,th9,th10,th11,th12,srm,dispm,sod,num1,num2,num3,num4,num5,msr,mdp,zhu}]  investigations. The pairing physics, described in Ref. [\onlinecite{msr}] that characterizes the pairing of slowly nucleated CFs in the intercorrelated superfluid state, with increasing distance, is an exceptionally good description of the superfluid phase at arbitrary $ d \lesssim l_B$. On the other hand, the fate of the superfluid phase at larger $d$ is not certain.

We point out that the monolayer and bilayer system possess a common pairing structure in the Son's formalism, and that the pairing of Ref. [\onlinecite{msr}] of two Fermi seas of CFs, as the natural outcome of the evolution of the superfluid state, is analogous to the PH Pfaffian pairing in the case of monolayer. Again we can argue that the opposite layer CF pairing wave function represents a critical state, which,  in the scope of a generalized Son's description, smoothly connects to an intermediate state that does not possess a Goldstone mode, but has an algebraic off-diagonal long-range order (ODLRO).

We discuss in the following section, Section II, the case of monolayer, and then, in Section III, the case of bilayer quantum Hall system. Conclusions are in Section IV.

\section{Monolayer}

We begin by considering a single Dirac fermion which was  proposed to effectively describe
half-filled LL of electrons [\onlinecite{son}], with $s$-wave pairing between spinor components.
We will neglect the presence of gauge fields in the following mean field treatment. The $s$-wave pairing suggested in Ref. [\onlinecite{son}], can be expressed by the  Bogoliubov - de Gennes Hamiltonian in the Nambu - Gorkov notation,
\begin{eqnarray}\label{BdGsD}
H &=&\frac{1}{2} \sum_{\bf k} \left[
  \begin{array}{cc}
   \Psi^\dagger ({\bf k}) & \tilde{\Psi}(-{\bf k}) \\
  \end{array}
\right]  \\ \nonumber
&\times&\left[
  \begin{array}{cc}
   {\cal D} ({\bf k}) & {\cal P}({\bf k}) \\
   {\cal P}^\dagger ({\bf k}) & - {\cal D} (-{\bf k}) \\
  \end{array}
\right]
\left[
  \begin{array}{c}
   \Psi ({\bf k}) \\
   \tilde{\Psi}^\dagger (-{\bf k}) \\
  \end{array}
\right],
\end{eqnarray}
where $\Psi ({\bf k})$ denotes a two-component spinor with momentum ${\bf k}$,
\begin{equation}
\Psi ({\bf k}) = \left[
  \begin{array}{c}
   \Psi_{a} ({\bf k}) \\
   \Psi_{b} ({\bf k}) \\
  \end{array}
\right], \,\,\,\,
\tilde{\Psi} ({\bf k}) = \left[
  \begin{array}{c}
   \Psi_{b} ({\bf k}) \\
   \Psi_{a} ({\bf k}) \\
  \end{array}
\right],
\end{equation}
and
\begin{equation}
{\cal D}({\bf k}) = \left[
  \begin{array}{cc}
   - \mu & k_x - i k_y \\
   k_x + i k_y & - \mu \\
  \end{array}
\right]=-\mu\sigma_0+k_x \sigma_x+k_y \sigma_y, \label{dmatrix}
\end{equation}
and  $2 \times 2$ matrix ${\cal P}({\bf k})$ describes Cooper pairing between $a$ and $b$ spinor components
\begin{equation}
{\cal P}({\bf k}) = \left[
  \begin{array}{cc}
   \Delta_s & 0\\
    0 & - \Delta_s \\
  \end{array}
\right]=\Delta_s \sigma_z  , \label{right_pairing_sD}
\end{equation}
or more explicitly
\begin{equation}
\delta {\cal H} = \sum_{\bf k} \{ - \Delta_s \Psi_{a}({\bf k}) \Psi_{b}(-{\bf k}) +  h.c. \}. \label{rightH}
\end{equation}
Here, $\sigma_0$ is the $2\times2$ identity matrix, while $\sigma_x$ and $\sigma_y$ are the standard Pauli matrices. Throughout the paper we set $\hbar = 1$, and the Fermi velocity, $v_F = 1$.  $\mu$ denotes a chemical potential equal to $\mu = \sqrt{B} = k_F$, where $B$ and $k_F$ are the external magnetic field and Fermi
vector, respectively. The dispersion of  Bogoliubons has
 the rotationally symmetric form, $
E_k^2 = (k \pm \mu)^2 + \Delta_{s}^2,$
where $k\equiv |{\bf k}|$.
This construction is considered in the literature
as a basis for a PH symmetric Pfaffian system.

However,  a different type of pairing is also possible with the pairing matrix
\begin{equation}
{\cal P}({\bf k}) = \left[
  \begin{array}{cc}
   0  & \alpha k_x \\
   -\alpha k_x  & 0 \\
  \end{array}
\right], \label{second_pairingx}
\end{equation}
or more explicitly
\begin{eqnarray}
\delta {\cal H}' =\sum_{{\bf k}} \alpha k_x  \{ \Psi_{a}^\dagger ({\bf k}) \Psi_{a}^\dagger(-{\bf k}) - \Psi_{b}^\dagger({\bf k}) \Psi_{b}^\dagger(-{\bf k}) \} + h.c. \nonumber \\
\label{anisotropicHx}
\end{eqnarray}
where $\alpha$ is a  constant, or
\begin{equation}
{\cal P}({\bf k}) = \left[
  \begin{array}{cc}
   0  &  \beta k_y\\
   \beta k_y & 0 \\
  \end{array}
\right], \label{second_pairing_sy}
\end{equation}
or more explicitly
\begin{eqnarray}
\delta {\cal H}'= \sum_{\bf k}  \beta k_y  \{ \Psi_{a}^\dagger({\bf k}) \Psi_{a}^\dagger(-{\bf k}) + \Psi_{b}^\dagger({\bf k}) \Psi_{b}^\dagger(-{\bf k}) \} + h.c.  \nonumber \\
\label{anisotropicHy}
\end{eqnarray}
where  $\beta$ is a constant. These two pairing possibilities, which we may associate with $p$-wave or triplet pairing among spinor components, by themselves make anisotropic gapless systems with two Dirac cones at two Fermi points at
$(k_x = 0, k_y = \pm \mu )$, and $(k_x = \pm \mu, k_y = 0 )$, respectively. (Other $p$-wave states  that respect PH symmetry are possible, as explained in Ref. [\onlinecite{mdcj}], but they would have more Fermi points and thus are not likely candidates with respect to the gain
in ground state energy due to pairing.)

Note that the pairings in (\ref{rightH}), (\ref{anisotropicHx}), and (\ref{anisotropicHy}) are invariant under the effective particle-hole transformations (of the underlying electron system), as explained in Ref. [\onlinecite{son}], up to a gauge transformation.

To further understand the pairings, we now consider the chirality operator $\frac{{\vec{\sigma}} \cdot{\vec k}}{k}$, and its eigenstates
\begin{equation}
|+ \rangle = \frac{1}{\sqrt{2}} \left[
  \begin{array}{c}
   1 \\
   \frac{k_+}{k} \\
  \end{array}
\right], \,\,\,\,
|- \rangle = \frac{1}{\sqrt{2}} \left[
  \begin{array}{c}
   - 1 \\
   \frac{k_+}{k} \\
  \end{array}
\right].\label{chiral-eigenstates}
\end{equation}
We can  introduce Dirac operators with a definite chirality
\begin{equation}
\Psi_+ ({\bf k}) = \frac{1}{\sqrt{2}}( \Psi_a ({\bf k}) +  \frac{k_-}{k} \Psi_b ({\bf k})) , \label{+ch}
\end{equation}
and
\begin{equation}
\Psi_- ({\bf k}) = \frac{1}{\sqrt{2}}( - \Psi_a ({\bf k}) +  \frac{k_-}{k} \Psi_b ({\bf k})) , \label{-ch}
\end{equation}
to find that
\begin{equation}
\Psi_a ({\bf k})  \Psi_b ( - {\bf k})
=- \frac{1}{2}  \frac{k_+}{k} [\Psi_+ ({\bf k})  \Psi_+ ( - {\bf k}) +   \Psi_- ({\bf k}) \Psi_- (-{\bf k})],
\label{s_into_p}
\end{equation}
with $k_\pm\equiv k_x\pm i k_y$.
We can clearly see from Eq.\ (\ref{s_into_p}) that in the chirality basis i.e. the eigenbasis of the non-interacting system, the pairing (\ref{rightH}), in fact, describes a pairing in the odd ($p$-wave) channel, with $p_+$, as a characteristic chirality of the PH Pfaffian.

On the other hand, model wave function constructions for the PH Pfaffian in the lowest LL (LLL) subspace may be expressed in different ways, but they always include Pfaffian pairing function,
\begin{eqnarray}
Pf\{ \frac{1}{(z_i^* - z_j^*)} \} \sim \nonumber \\
 \sum_P {\rm sgn} \; P \prod_{i=1}^{N/2} \frac{1}{(z_{P(2 i - 1)}^* - z_{P(2 i)}^*)},  \label{cPf}
\end{eqnarray}
where the sum is over all permutations $(P)$ of $N$ objects, and we made an assumption
that the uniform external magnetic field is defined as $\vec{B} = - B \; \hat{e}_z, B > 0$. Here $z_i$ denotes the complex (2D) coordinate of the $i^{th}$ electron,  $i = 1, \ldots , N$. Thus Cooper pair wave function i.e. pairing function, $ g(r_{ij}) \sim \frac{1}{(z_i^* - z_j^*)}$, decays with distance $r_{ij} = |z_i - z_j|$, as one would expect in a BCS theory. (The complex conjugation of $z$'s is due to the opposite chirality, built in PH Pfaffian state,  with respect to the
direction defined by the external magnetic field.) The form of the pairing function, $ g(r_{ij}) \sim \frac{1}{(z_i^* - z_j^*)}$, is expected not just because of the usual pairing behavior but, if we use  it as a part of a negative flux insertion [\onlinecite{tj}], as shown in Ref. [\onlinecite{mvmtj}], we can easily analytically generate and reproduce the edge states expected of the PH Pfaffian QH phase: charge edge mode plus Majorana neutral edge mode in the opposite direction. This assertion is  true even when we consider the necessary projection to the LLL of the model wave function with the anti-holomorphic part described by the function in (\ref{cPf}). In the long distance limit i.e. when $k << 1/l_B$, where $l_B = 1/\sqrt{B}$ is the magnetic length, we may neglect the projection due to effectively commuting variables, $z$ and $z^*$, in LLL.

Based on the proceeding arguments for the form of the pairing wave function we expect that in the Son's formalism we should consider extended $s$-wave pairing i.e. $ \lim_{k \rightarrow 0} \Delta_s \sim |{\bf k}|\equiv k$. Namely, the effective description of the Hamiltonian,  Eqs. (\ref{BdGsD}-\ref{rightH}), in the chirality basis, (\ref{+ch}) and
(\ref{-ch}), is
\begin{eqnarray}
H&=& \sum_{\bf k} (k - \mu) \Psi_+^\dagger ({\bf k}) \Psi_+ ({\bf k}) +  \nonumber \\
  && \sum_{\bf k} (- k - \mu) \Psi_-^\dagger ({\bf k}) \Psi_- ({\bf k}) +  \nonumber \\
&& \sum_{\bf k} \{\frac{1}{2} \frac{k_+}{k} \Delta_s \Psi_+ ({\bf k})  \Psi_+ ( - {\bf k}) + h.c.\} + \nonumber \\
&& \sum_{\bf k} \{\frac{1}{2} \frac{k_+}{k} \Delta_s \Psi_- ({\bf k})  \Psi_- ( - {\bf k}) + h.c.\}. \label{pH}
\end{eqnarray}
Taking $ \lim_{k \rightarrow 0} \Delta_s \sim |{\bf k}|\equiv k$, the positive chirality ($\Psi_+ ({\bf k})$) part (as well the negative chirality part) constitutes the usual $p$-wave description as given in Ref. [\onlinecite{rg}], and this ensures the weak pairing with $ g(z) \sim \frac{1}{z}$ at long distances $(|z| \rightarrow \infty)$, in the physically relevant - the positive chirality sector. Namely, in our case, compare with the notation of Ref. [\onlinecite{rg}], the Fourier transform of the pairing function, $g(\bf r)$, can be expressed as
\begin{equation}
g_{\bf k} = \frac{v_{\bf k}}{u_{\bf k}} = \frac{-(E_{\bf k} - \xi_{\bf k})}{\Delta_{\bf k}},
\end{equation}
where $ \xi_{\bf k} = k - \mu$, $\Delta_{\bf k} = \frac{k_+}{k} \Delta_s$, and $E_{\bf k}^2 = \xi_{\bf k}^2 + |\Delta_{\bf k}|^2$. The minimum of $E_{\bf k}$ is at the Fermi momentum, $k = \mu$, but we are interested in the long distance behavior with momenta, $ k \sim 0$. Thus
$ \lim_{k \rightarrow 0} g_{\bf k} \sim 1/k_+$, and $ \lim_{r \rightarrow \infty} g_{\bf r} \sim 1/z$, if
$ \lim_{k \rightarrow 0} \Delta_s \sim k $.
Here we should note that in the Son's theory the natural choice for the direction of the magnetic field is $\vec{B} =  B \; \hat{e}_z, B > 0$ (or the choice of the coordinate system). This is natural because the density of the positive chirality particles is proportional to the flux density, $ \bar{\Psi} \gamma_0 \Psi = \vec{\partial} \times \vec{A} = B/(2 \pi) > 0$ (see Eq.(24) in Ref. [\onlinecite{sonr}]). Thus we would have in the Son's formalism $ g(z) \sim \frac{1}{z}$ instead of the usual $ g(z) \sim \frac{1}{z^*}$ due to the different convention for the direction of the external field.

On the other hand, taking $ \lim_{k \rightarrow 0} \Delta_s$ to be a constant, would lead again to the pairing at the Fermi surface (circle) and the Cooper pair wave function would behave as $ g(z) \sim \frac{1}{z |z|}$ at long distances. This certainly would not lead to a ``nice" expression for a quantum Hall wave function, but more importantly, we do not have analytical means to derive the edge states (i.e. physically motivated low energy states that constitute subspace with charge - chiral boson and Majorana mode) that we can associate with the PH Paffian QH state, in this case. The quantum Hall state consists of (entangled) neutral (pairing)  and charge parts, and the underlying topological order has to be claimed for that construct. We note that this long-distance, universal pairing behavior coincides with the behavior of the critical state between strong and weak $p$-wave pairing  as demonstrated in Ref. [\onlinecite{rg}].

But the analysis of edge states in Ref. [\onlinecite{mvmtj}] based on the construction in (\ref{cPf}) still does not guarantee that in the LLL there exists a model interaction for which these edge states would make a zero energy subspace (or an interaction that would delineate this subspace with respect to higher energy bulk excitations), and stabilize a PH Pfaffian phase. In connection with this, we showed that, under very
natural assumptions for the constructions of model wave functions, we can conclude  that the order parameter for PH Pfaffian is nonanalytic in the neighborhood of the $\vec{k} = 0$ expansion point (if we are in the two-component Dirac formalism). Thus a gradient (Landau-Ginzburg) expansion around
$\vec{k} = 0$ point is not well defined. This implies that, because the Dirac spinor description is at the foundation of the PH symmetric description, PH Pfaffian, in the PH symmetric half-filled LL, is a critical - gapless  state, and does not represent a stable phase of topological matter. This is consistent with numerical [\onlinecite{pjds,dh,ph,pp}], and analytical work [\onlinecite{wc}]. The numerical experiments of these references are consistent with a ``Schroedinger cat" superposition of Pfaffian and anti-Pfaffian, in the presence of PH symmetry, but we find that a careful assessment of the role of PH Pfaffian in the critical region between Pfaffian and anti-Pfaffian phases [\onlinecite{dh}] is still missing.

We may still pose the question regarding the role of triplet pairings in (\ref{anisotropicHx}) and (\ref{anisotropicHy}).
Expressed in the chirality basis the superpositions in (\ref{anisotropicHx}) and (\ref{anisotropicHy})  both contain a $p$-wave which chirality corresponds to the PH Pfaffian, and
symmetric superposition of Pfaffian and anti-Pfaffian. Namely,
\begin{eqnarray}
k_x (\Psi_a ({\bf k})  \Psi_a ( - {\bf k}) - \Psi_b ({\bf k})  \Psi_b ( - {\bf k})) = \nonumber \\
\frac{1}{4 } (2 k_+ + k_- + \frac{(k_+)^3}{|k|^2})
 \Psi_+ ({\bf k})  \Psi_+ ( - {\bf k}) + ... ,
\label{t_into_pf2}
\end{eqnarray}
and,
\begin{eqnarray}
k_y (\Psi_a ({\bf k})  \Psi_a ( - {\bf k}) + \Psi_b ({\bf k})  \Psi_b ( - {\bf k})) = \nonumber \\
\frac{1}{4 i} (2 k_+ - k_- - \frac{(k_+)^3}{|k|^2})
 \Psi_+ ({\bf k})  \Psi_+ ( - {\bf k}) + ... ,
\label{t_into_pf1}
\end{eqnarray}
where, as before,  $\Psi_+ ({\bf k})$ denote positive energy (particle) solutions, and missing terms have negative  energy (hole i.e. higher in energy) contributions. The symmetric superposition is present in this basis as $k_- + \frac{(k_+)^3}{|k|^2}$ i.e. as a superposition of Pfaffian ($k_-$) and anti-Pfaffian ($\frac{(k_+)^3}{|k|^2}$), while $k_+$ represents a PH Pfaffian component.

If we consider a BdG Hamiltonian with both $s$-wave and $p$-wave pairing present, in the explicitly symmetric PH case, we have the following expression for the low energy projected part of the pairing  Hamiltonian,
\begin{eqnarray}
\{ -  \frac{\Delta_s}{2}  \frac{k_+}{k}  + \frac{\alpha}{8} (2 k_+ + k_- + \frac{(k_+)^3}{|k|^2})\}
 \Psi_+ ({\bf k})  \Psi_+ ( - {\bf k}) +  ..., \nonumber \\
\label{s_and_pProj}
\end{eqnarray}
if the $p$-wave is described by (\ref{anisotropicHx}), and a similar expression we would have if the $p$-wave is given by (\ref{anisotropicHy}). The analytical (RPA) considerations in Ref. [\onlinecite{wc}] preclude any pairing in the $k_+$ channel, and thus, with the assumed $ \lim_{k \rightarrow 0} \Delta_s \sim k$  behavior, we may envision a cancellation of the singlet and triplet spinor component pairing, which would leave the anisotropic combinations in Cooper pairing, $k_- \pm \frac{(k_+)^3}{|k|^2}$, as viable pairing instabilities. The numerical work is not equivocal at this point: While the Refs. [\onlinecite{pjds,dh,ph,pp}] suggest that superpositions of Pfaffian and anti-Pfaffian states are relevant in the half-filled LL, Ref. [\onlinecite{jp}] suggests that maybe the doubling of the expected ground state degeneracy, equal to $ 12 = 2 \times 6$, is not due to the symmetric and antisymmetric superpositions of Pfaffian and anti-Pfaffian states, but due to the two possibilities for anisotropic Cooper pairs. Also Ref. [\onlinecite{wy}] argues for an anisotropic superposition of Pfaffian and anti-Pfaffian. A small, parallel to the plane magnetic field may induce this scenario as demonstrated experimentally in Refs. [\onlinecite{l1,l2}].

Without an anisotropic agent,
it seems that the higher angular momentum expansion of the possible pairings instabilities in the scope of the Son's formalism with Cooper pair anisotropy scenario is not likely outcome (because of the inclusion of higher momenta in the underlying Lagrangian density expansion) consistent with Refs. [\onlinecite{dh,ph,pp}]. Nevertheless, the triplet pairings assume major role when we spoil PH symmetry by adding a mass term to the Dirac CF description of a half-filled LL. The mass term acts as some kind of a Zeeman energy term that will favor one or the other spinor component. We expect, based on   (\ref{anisotropicHx}) and (\ref{anisotropicHy}), depending on the sign of the mass term, either PH Pfaffian or Pfaffian to constitute the ground state. In Appendix B of Ref. [\onlinecite{mdcj}]  such a scenario is described assuming, in addition to (\ref{anisotropicHx}) (or (\ref{anisotropicHy}))  the presence of the usual $s$-wave pairing i.e. for which $ \lim_{k \rightarrow 0} \Delta_s$ is a constant, and the  PH symmetry breaking mass term. Thus the presence of PH Pfaffian (paradoxically) for sufficiently strong  PH breaking, which is consistent with experiments [\onlinecite{zf}] (see also recent Ref. [\onlinecite{ch}]), can be explained on the basis of the Dirac CF formalism.

\section{Bilayer}

The bilayer system that we will discuss consists of two half-filled LL layers at distance $d$ between layers. The most
important feature of this system, a correlated interlayer excitonic phase at small distances between layers, $ d \lesssim l_B$, was predicted and discovered experimentally [\onlinecite{eis}]. Due to interlayer correlations and underlying bosonic ODLRO the system at small distances ($ d \lesssim l_B$) possesses a Goldstone mode [\onlinecite{wz,moon}].

Our understanding of the system at small $d$ is very much based on a model wave function, so-called (111) state,
\begin{eqnarray}
 \prod (z_{i \uparrow} - z_{j \uparrow}) \prod (z_{k \downarrow} - z_{l \downarrow})\prod (z_{m \uparrow} - z_{n \downarrow}) ,
 \label{111}
\end{eqnarray}
for the intercorrelated state. We omitted Gaussian factors, and  $z_{\uparrow}$'s and $z_{\downarrow}$'s represent complex coordinates of two kinds of electrons in 2D plane. For decoupled layers, at $ d \rightarrow \infty$, in the LLL, we have two decoupled CF Fermi-liquid-like condensates, which may be described by the Son's formalism. The model wave function is \begin{eqnarray}
\Psi_{FL}({\bf r}_\uparrow) \prod (z_{i \uparrow} - z_{j \uparrow})^2   \Psi_{FL}({\bf r}_\downarrow)   \prod (z_{k \downarrow} - z_{l \downarrow})^2, \nonumber \\
\label{2FL}
\end{eqnarray}
i.e. a product of two (unprojected to the LLL) Rezayi-Read wave functions [\onlinecite{rr}]. We again omitted Gaussian factors, and $\Psi_{FL}({\bf r}_\sigma)$, $\sigma = \uparrow,\downarrow$ denote two Fermi seas - Slater determinants of plane (free) waves. Each Rezayi-Read wave function, when projected to the LLL, describes the half-filled systems [\onlinecite{rh}].

These two extremes, described by Eqs. (\ref{111}) and (\ref{2FL}), were used in Ref. [\onlinecite{srm}] to suggest a mixed state representation, with both inter (\ref{111}) and intra (\ref{2FL}) correlations as a good interpolating ansatz for the system at finite distances. Building on this proposal, an understanding of the evolution of the excitonic state with distance was achieved in Ref. [\onlinecite{msr}]. Namely the part in the mixed state proposal with intracorrelations and natural CF representations has to be modified by a $p$-wave pairing among different layer CFs, in order to extremely well describe the QH superfluid i.e. excitonic state at finite distance, $ d \lesssim l_B$. The necessity for the special chirality $p$-wave pairing was also recognized in
Ref. [\onlinecite{pm}], as a way to describe superfluid evolution (disordering) in order to capture the most basic disordering due to the zero point motion of the system.

Following the proposal of Ref. [\onlinecite{msr}], in Ref. [\onlinecite{mdp}], a model wave function as a natural outcome of the superfluid disordering was proposed:
\begin{eqnarray}
\Psi_{c}& = & \prod (z_{i \uparrow} - z_{j \uparrow})^2 \times \prod (z_{k \downarrow} - z_{l \downarrow} )^2 \nonumber \\
&& {\rm Det} \{\frac{1}{(z_{m \uparrow}^{*} - z_{n \downarrow}^{*})}\}, \label{ppairing}
\end{eqnarray}
for long distance behavior. Here
\begin{eqnarray}
{\rm Det} \{ \frac{1}{(z_{m \uparrow}^{*} - z_{n \downarrow}^{*})} \} \sim \nonumber \\
 \sum_\sigma {\rm sgn} \; \sigma \prod_{i=1}^{N/2} \frac{1}{(z_{i \uparrow}^* - z_{\sigma(i) \downarrow}^*)},  \label{det}
\end{eqnarray}
where the sum is over all permutations ($\sigma$) of $N/2$ objects.
Thus, at the end of the disordering all (different layer) CFs are paired in the way of $p$-wave that is of the opposite chirality with respect to the one induced by the external magnetic field [\onlinecite{msr,pm}]. Analogously to the previously discussed PH Pfaffian model wave function, the right chirality, $\frac{|z|}{z^*}$, was combined with a natural decay ($\frac{1}{|z|}$) function.

We may use the Cauchy identity,
\begin{eqnarray}
{\rm Det} \{ \frac{1}{(z_{m \uparrow} - z_{n \downarrow})} \}  \sim
\frac{\prod (z_{i \uparrow} - z_{j \uparrow} ) \times \prod (z_{p \downarrow} - z_{q \downarrow} )}{\prod (z_{k \uparrow} - z_{l \downarrow} )}, \nonumber \\
\end{eqnarray}
to rewrite  the wave function in the following form
\begin{eqnarray}
\Psi_{c}& = & \prod_{i<j} (z_m - z_n ) \times \nonumber \\
&&  \frac{\prod |z_{i \uparrow} - z_{j \uparrow} |^2 \times \prod |z_{p \downarrow} - z_{q \downarrow} |^2}{\prod |z_{k \uparrow} - z_{l \downarrow}|^2}. \nonumber \\
&&
\label{twodet}
\end{eqnarray}
By looking at this model wave function and assuming that the system is in $\nu =1$ integer quantum Hall effect (IQHE) phase (for charge degrees of freedom) the most natural quasiparticle representation seems to be composite boson (CB) representation, where CBs, due to the presence of the additional factor next to the (111) state are ``disordered bosons": bosons interacting with long distance interactions or gauge fields [\onlinecite{mdp}].

CB description is not as well founded as the CF description, and at this point the intricate picture of bosonic disordering is not complete, but it may be used to motivate the appearance of various low-lying states in the intermediate region [\onlinecite{mdp}]. It is not clear whether the wave function in Eq. (\ref{twodet}) is a relevant critical state (model wave function for an intermediate phase or a transition point), or a function in the universality class of the (111) state [\onlinecite{sod}]. (The latter would imply the same (111) phase for any finite distance between the layers, if the $p$-wave pairing between different layer CFs is present.) We expect, based also on the preceding discussion concerning monolayer PH Pfaffian that due to the non-analytic behavior around $\vec{k} = 0$ in the Son's formulation, the state  in Eq. (\ref{twodet}) is a critical state, if the underlying Hamiltonian is strictly PH symmetric. To further explain the analogy we recapitulate the CF pairing bilayer physics using the Son's formalism.

To understand the relevant pairings in the quantum Hall bilayer at the mean field level, we can go back to the Bogoliubov - de Genes Hamiltonian in the monolayer case, Eqs. (\ref{BdGsD}) and (\ref{dmatrix}), and assign layer indexes,$\uparrow$ and $\downarrow$, by the following substitutions,
\begin{eqnarray}
\Psi ({\bf k}) = \left[
  \begin{array}{c}
   \Psi_{a} ({\bf k}) \\
   \Psi_{b} ({\bf k}) \\
  \end{array}
\right] \rightarrow
 \left[
  \begin{array}{c}
   \Psi_{a \uparrow} ({\bf k}) \\
   \Psi_{b \uparrow} ({\bf k}) \\
  \end{array}
\right] \nonumber \\
\tilde{\Psi} ({\bf k}) = \left[
  \begin{array}{c}
   \Psi_{b} ({\bf k}) \\
   \Psi_{a} ({\bf k}) \\
  \end{array}
\right] \rightarrow
\left[
  \begin{array}{c}
   \Psi_{b \downarrow} ({\bf k}) \\
   \Psi_{a \downarrow} ({\bf k}) \\
  \end{array}
\right].
\end{eqnarray}
The ensuing pairing contributions, with respect to pairing matrices in (\ref{right_pairing_sD}), (\ref{second_pairingx}), and (\ref{second_pairing_sy}), in the monolayer case, are
\begin{equation}
\delta {\cal H}_b = \sum_{\bf k} \{ \Delta_s^b \Psi_{a \uparrow}^\dagger ({\bf k}) \Psi_{b \downarrow}^\dagger (-{\bf k}) +  h.c. \}. \label{swbH}
\end{equation}
and
\begin{eqnarray}
\delta {\cal H}'_b \sim \;\;\;\;\;\;\;\;\;\;\;\;\;\;\;\;\;\;\;\;\;\;\;\;\;\;\;\;\;\;\;\;\;\;\;\;\;\;\;\;\;\;\;\;\;\;\;\;\;\;\;\;\;\;\;\;\;\;\;\; \nonumber \\
\sum_{{\bf k}} k_{x(y)}  \{ \Psi_{a \uparrow}^\dagger ({\bf k}) \Psi_{a \downarrow}^\dagger(-{\bf k}) \mp \Psi_{b \uparrow}^\dagger({\bf k}) \Psi_{b \downarrow}^\dagger(-{\bf k}) \} + h.c. \nonumber \\
\label{pwbH}
\end{eqnarray}
These are triplet pairings with respect to the layer index. The $s$-wave pairing in (\ref{swbH}) describes the ``anti-chiral" channel  - $p$ - wave in the opposite sense with respect to the induced chirality of the external magnetic field (perpendicular to the plane). An analysis similar to the one in the case of monolayer, leads to the conclusion that the description of the CF state in Eq. (\ref{ppairing}) in the Son's formalism has to assume $ \lim_{k \rightarrow 0} \Delta_s^b \sim |\vec{k}|\equiv k$, and  this describes a critical state.

The most recent numerical results in Ref. [\onlinecite{zhu}] for the intermediate region motivate the following mixed state construction as described in Ref. [\onlinecite{dispm}] ($\Psi_2$ in the notation of the same reference):
\begin{eqnarray}
&&{\cal A} \{ \prod (z_{i \uparrow} - z_{j \uparrow}) \prod (z_{k \downarrow} - z_{l \downarrow})\prod (z_{m \uparrow} - z_{n \downarrow}) \nonumber \\
&& \prod (z_{r \uparrow} - w_{s \uparrow})^2 \prod (z_{p \downarrow} - w_{q \downarrow})^2 \nonumber \\
&& \Psi_{FL}({\bf r}_\uparrow) \prod (w_{t \uparrow} - w_{u \uparrow})^2   \Psi_{FL}({\bf r}_\downarrow)   \prod (w_{f \downarrow} - w_{h \downarrow})^2 \} , \nonumber \\
&& \label{mix}
\end{eqnarray}
where ${\cal A}$ is  an overall antisymmetrization. The part with (111) correlations may be followed with the Jastrow-Laughlin factors as in Eq. (\ref{twodet}), but that will not change main conclusions reached in Ref. [\onlinecite{dispm}] (in the scope of a Chern-Simons (CS) description) for an intermediate phase: The pseudospin mode, which was a Goldstone mode in the (111) phase, is gapped and the phase possesses algebraic ODLRO with the exponent that depends on the ratio between densities of CBs and CFs. The disappearance of the Goldstone mode directly correlates with the results of Ref. [\onlinecite{zhu}] (see also Ref. [\onlinecite{mdp}]).

Thus the description given in Ref. [\onlinecite{dispm}] captures the main characterization of the intermediate phase. Still there is an interesting even-odd effect, as detected in Ref. [\onlinecite{zhu}], that for an odd number of electrons in each layer the pseudospin excitation is without a gap. It was noted in Ref. [\onlinecite{zhu}] that this may be a consequence of intralayer pairing. The most probable pairing would be of Pfaffian kind, and this may be occur in the CF part (in the two CF Fermi seas) of the model wave function in Eq. (\ref{mix}). The arguments of Ref. [\onlinecite{dispm}] can be easily extended to this case by applying the parton modeling of the CF part - an electron consisting of slaved charged boson and neutral (composite) fermion that will pair. Assuming the low-momentum decoupling of slave boson and neutral fermion we can (following Ref. [\onlinecite{dispm}]) recalculate the exponent of algebraic ODLRO and find that it is equal to $ \sqrt{n_f}/(\sqrt{n_f} + \sqrt{n_b})$, where $ n_f$, and $n_b$ are CF, and CB densities, respectively. An assumption is made that $n_b < n_f$.

This mixed state phenomenological approach has a support in the recent description of an excitonic metal in Ref. [\onlinecite{bar}] and its experimental detection in Ref. [\onlinecite{zib}]. Though in our case there is no seemingly natural distinction of CFs and CBs (for fixed layer index) as in the physics of the excitonic metal underlined by distinguishable degrees of freedom - compare the model wave function in Ref. [\onlinecite{bar}] and the one in Eq. (\ref{mix}), the experimental detection in a real system in Ref. [\onlinecite{zib}] implies that in the model wave function in in Ref. [\onlinecite{bar}] an overall antisymmetrization is assumed and present (due to the indistinguishability of real electrons), but this can be neglected in a theoretical model. Similarly, in our case, two kinds of coexisting condensates can be treated as if two  kinds of electrons are present in each layer.

There is a growing body of theoretical evidence that the Halperin Lee Read (HLR) theory [\onlinecite{hlr}]  and Son's formalism for CFs at filling one-half do not differ in important physical characterization, but still, it is interesting and instructive to apply the Son's formalism to CFs in the mixed state in Eq. (\ref{mix}).
We propose the following Lagrangian density for the description at intermediate distances between the layers of the  bilayer (neglecting the possibility for intra-pairing):

\begin{eqnarray}
&&{\cal L} = \;\;\;\;\;\;\;\;\;\;\;\;\;\;\; \nonumber \\
&& - \sum_{\sigma ,i} \frac{|(\partial_i - a_i^c - \sigma a_i^s + \sigma a_i^b) b_\sigma |^2}{2 m_b} +  \nonumber \\
&& \sum_\sigma b_\sigma^\dagger (\partial_0 - a_0^c - \sigma a_0^s + \sigma a_0^b) b_\sigma + \nonumber \\
&&   \frac{\epsilon_{\mu \nu \lambda}}{4 \pi} a_\mu^b \partial_\nu a^b_\lambda +                     \sum_{\sigma} V_b \; \rho_\sigma^b \rho_\sigma^b + \nonumber \\
&& \sum_\sigma \bar{\Psi}_\sigma \gamma^\nu (\partial_\nu - a_\nu^c - \sigma a_\nu^s) \Psi_\sigma + \nonumber \\
&& \frac{\epsilon_{\mu \nu \lambda}}{2 \pi} (a_\mu^c \partial_\nu A_\lambda + \frac{1}{2} A_\mu \partial_\nu A_\lambda) + \nonumber \\
&& + (\nabla \times \vec{a}_c)^2 V_{c}(r) + (\nabla \times \vec{a}_s)^2 V_{s}(r). \label{L}
\end{eqnarray}
Next to two copies of Son's Lagrangian for two layers, $\sigma = \uparrow, \downarrow$; and by introducing
two gauge fields, $a^\mu_\sigma , \sigma = \uparrow, \downarrow$; in the combinations
\begin{equation}
a^\mu_c = \frac{ a^\mu_\uparrow  + a^\mu_\downarrow}{2} \; \;{\rm and}\;\;a^\mu_s = \frac{ a^\mu_\uparrow  - a^\mu_\downarrow}{2},
\end{equation}
we have also introduced two, $b_\sigma, \sigma = \uparrow, \downarrow$, bosonic fields, with
densities, $\rho_\sigma^b = b_\sigma^* b_\sigma, \sigma = \uparrow, \downarrow$, and $V_b$ is a repulsive short-range interaction  among same layer composite bosons [\onlinecite{zh}], which are necessary to ensure short-distance exclusion among composite bosons, which represent (dressed) electrons. Thus next to ``fermionic vortices", described by $\Psi_\sigma$, see Ref. [\onlinecite{sonr}], we introduced ``bosonic vortices" (neutral composite bosons). Thus we generalized the constraint in Eq. (24) in Ref.  [\onlinecite{sonr}] to
\begin{equation}
\sum_{\sigma} b_{\sigma}^{\dagger} b_{\sigma} + \sum_{\sigma} \bar{\Psi}_{\sigma}\gamma_0 \Psi_{\sigma} =
\frac{B}{2 \pi}.
\end{equation}
By coupling to $a_s^\mu$ the bosonic fields, we keep fixed the difference (equal to zero) between the total number of vortices in each layer, thus conserving the number of vortices in each layer. By introducing gauge field $a_b^\mu$ we also
keep fixed the difference between the number of bosons in each layer. (The presence of $a_b^\mu$ is needed to ensure bosonic statistics for $b_{\sigma}$ fields.)

The most interesting conclusion that we can draw from this application of the Dirac CF formalism is that the bosonic part necessarily acquires the additional correlations next to the basic (111), as described by Eq. (\ref{twodet}). The ensuing presence of three gauge fields that couple to the bosonic fields may lead to fractional excitations in the low-energy sector, but not of quantized pseudospin. Thus we cannot expect deconfined meron eigenstates in the intermediate region and exact topological degeneracy.

The preceding description of the mixed state in the Son's formalism implies that the wave function in Eq. (\ref{twodet}) describes a critical state which may smoothly connect the (111) phase to the intermediate phase. The state may be smoothly continued in the intermediate region by the gradual inclusion of Dirac CFs. From the usual RPA treatment of the mixed state CS description in Ref. [\onlinecite{dispm}] (which we summarized here), we can conclude that the intermediate phase is incompressible in charge channel (IQHE), and does not possess a Goldstone mode in accordance with the results in Ref. [\onlinecite{zhu}].

The presence of interlayer (Coulomb) interaction in the system will necessarily break PH symmetry inside each layer. Also the density of (intracorrelated) CFs will gradually increase (with $d$, in the intermediate phase) in this effective
description, but it will be always less than the nominal, corresponding to half-filled LLs density of electrons in each layer. These conditions will be favorable for an establishment of intra-paired (of Pfaffian kind) state inside each layer; the reduced density will disfavor the slip into Fermi-liquid-like state of the weak-pairing state, inside each layer.

The task of the description of the intermediate phase, without the phenomenological division of electrons inside each layer, is desirable, and may start  by considering the effective Lagrangian for $s$-wave interlayer pairing in the Dirac CF formalism, as in Ref. [\onlinecite{sod}], to which an intralayer pairing is added. Formally, through the Anderson-Higgs mechanism, the Goldstone mode (in the pseudospin channel) will become gapped. If we put aside the question of whether we should consider triplet channels in the Son's formalism, due to PH symmetry breaking inside layers and the preceding discussion concerning the $p$-wave pairing, a more difficult question is whether inter and intra pairing may coexist, or whether, given that we have the relevant wave functions (Eq.(\ref{mix})), an effective field theoretical description (other than in Eq.(\ref{L})) may exist.

We may begin the search for the effective theory by considering the effective description in the neutral channel (by neglecting or decoupling charge fluctuations inside each layer) and considering classical (not Dirac) CFs in the scope of a mean field treatment. We consider classical (HLR) fermions as we expect that at intermediate distances the PH symmetry inside each layer will be broken (we assume something that would require much more work to be captured in the Son's formalism). Any mean-field treatment will not be able to capture complex correlations of mixed states, but it is still interesting to see what a mean field treatment of inter and intra pairings can give or imply. The BCS Hamiltonian for the pairing of CFs is
\begin{eqnarray}
 H_{\rm eff} &= & \sum_{{\bf k}, \sigma} \xi_{\bf k} \Psi^\dagger_\sigma ({\bf k}) \Psi_\sigma ({\bf k}) + \nonumber \\
&& \sum_{{\bf k}, \sigma} \frac{1}{2} (\Delta_{\bf k} \Psi^\dagger_\sigma ({\bf k}) \Psi^\dagger_\sigma ({- \bf k}) + h.c.) + \nonumber \\
&& \sum_{{\bf k}, \sigma} \frac{1}{2} (\delta_{\bf k} \Psi^\dagger_\sigma ({\bf k}) \Psi^\dagger_{-\sigma} ({- \bf k}) + h.c.),
\end{eqnarray}
where $\xi_{\bf k} = \epsilon_{\bf k} - \mu , \epsilon_{\bf k} \sim {\bf k}^2 $, and $\Delta_{\bf k}$ and $\delta_{\bf k }$ represent intra and inter $p$-wave pairing function, respectively. Assuming generic BCS Hamiltonian that includes (constant) term : $ \sum_{{\bf k}} (g_a |\Delta_{\bf k}|^2 + g_e |\delta_{\bf k}|^2)$, we can conclude, after a straightforward analysis of free energy in this case, that the coexistence of pairings is impossible. It is only possible, in a special case when $ \Delta_{\bf k} = \Delta k_-$ and $ \delta_{\bf k} = \delta k_-$ (or $ \Delta_{\bf k} = \Delta k_+$ and $ \delta_{\bf k} = \delta k_+$), and, moreover, $\Delta = \pm \delta$. Thus we have two degenerate solutions, in which a symmetric (or antisymmetric) combination of CFs is in a Fermi-liquid-like state, and the remaining antisymmetric (symmetric) combination is in a $p$-wave state.

The $p$-wave is expected to have the chirality opposite to the one dictated by external field (based on a continuity
argument that takes into account smaller distance-$d$  behavior as we already discussed), and thus the intra-pairing has to be of a PH Pfaffian kind. The two states can not represent the intermediate phase in its generality, but they may be relevant states for the description of the putative phase transition between the intermediate and the Fermi-liquid-like phase present at large distances between the layers; their lower Bogoliubov bands describe a Fermi liquid behavior in one of the two superpositions of layer degrees of freedom, but also their upper Bogolibov  bands describe a  $p$-wave, likely critical behavior in the other (orthogonal) superposition. The $p$-wave behavior is likely critical for the same reason we discussed in the monolayer case; in the strict PH symmetric circumstances (for a system as a whole) the $p$-wave that respects this symmetry must be critical.

We can conclude that the mean-field treatment cannot capture the complex physics of the intermediate phase, but it is suggestive that the paired state in Eq. (\ref{twodet}) (at a smaller $d$), and the two states with both Fermi-liquid physics and PH Pfaffian pairing of symmetric and antisymmetric superpositions of layer degrees of freedom i.e. CFs
(at a larger $d$), may mark the boundaries of the intermediate phase.

\section{Conclusions}

In this work we explained why a PH Pfaffian state in the (PH symmetric) half-filled LLL of a monolayer, and an analogous state in the PH symmetric bilayer (in which each layer is half-filled LL) can be considered as critical states i.e. states that cannot describe a phase under PH symmetry. This is consistent with numerical work in Refs. [\onlinecite{pjds,dh,ph,pp}], in the case of monolayer, and, the most recent results,  in Ref. [\onlinecite{zhu}], in the case of a bilayer. We showed in the case of monolayer that the inclusion of a PH symmetry breaking (like LL mixing) may stabilize PH Pfaffian consistent with experiments [\onlinecite{zf}]. We expect that an inclusion of PH symmetry breaking in the bilayer will stabilize an analogous (opposite chirality $p$-wave pairing) state i.e.,  just as in the monolayer case, we can consider analytical pairings, which under PH symmetry breaking mass(es) may be stabilized. On the sphere, by choosing the PH symmetric shift we can stabilize the (111) excitonic or critical state for any distance between the layers [\onlinecite{msr,sod,ms}].
Nevertheless, on a torus, with no bias as shift on sphere, the evolution of the bilayer includes other phases that do not possess a Goldstone mode or behave as a CF Fermi-liquid-like phase [\onlinecite{zhu}], because they are stable phases in the presence of the underlying PH symmetry of the Hamiltonian
in the half-filled LLs.

In the monolayer case we find that the PH Pfaffian (as a gapped topological phase) cannot exist in a PH symmetric half-filled LL. We reached this conclusion by examining the intrinsic $s$-wave order parameter in the Son's formulation; analytic and non-analytic version lead to critical states i.e. gapless states (that cannot describe a gapped topological state). In the bilayer case, we cannot reach such a conclusion that eliminates any other scenario. If we assume an analytic, intrinsic $s$-wave pairing (which would lead to pairing function
 $ g(z) \sim \frac{1}{z^* |z|}$, instead of  $ g(z) \sim \frac{1}{z^*}$  in Eq. (\ref{ppairing})) this may be still a viable gapless state (with the right chirality [\onlinecite{msr}] but different decay function) - a representative of a gapless phase in the bilayer case. This CF pairing state may be in the same universality class of the (111) state, and the excitonic order may exist for any distance between layers [\onlinecite{sod}]. The CF representation invariably favors CF pairing in the bilayer [\onlinecite{th2,sod}]. Still this pairing  may be rather unstable, and give way to two Fermi-liquid-like states at large distances [\onlinecite{zhu}], and the intermediate phase as described in Ref. [\onlinecite{zhu}] and here.
\\
\\

\begin{acknowledgments}
The author would like to thank N. Regnault and I. Sodemann for valuable comments.
This research was supported in part by the National Science Foundation under Grant No. NSF PHY11-25915. The research was also supported by
the Ministry of Education, Science, and Technological
Development of the Republic of Serbia under project
ON171017.
\end{acknowledgments}

\end{document}